\documentclass[manuscript]{aastex}
\include{epsf}

\shorttitle{INTEGRATED BROADBAND COLORS}
\shortauthors{LEE, LEE \& GIBSON}

\begin{document}

\title{HORIZONTAL$-$BRANCH MORPHOLOGY AND\\
       THE PHOTOMETRIC EVOLUTION OF OLD STELLAR POPULATIONS}

\author{Hyun-chul Lee\altaffilmark{1}, Young-Wook Lee}
\affil{Center for Space Astrophysics, Yonsei University, Shinchon 134, 
    Seoul 120-749, Korea}

\and

\author{Brad K. Gibson}
\affil{Centre for Astrophysics \& Supercomputing, Swinburne University,
    Mail\#31, P.O.Box218, Hawthorn, Vic 3122, Australia}

\altaffiltext{1}{present address: Centre for Astrophysics \& Supercomputing, 
    Swinburne University, Mail\#31, P.O.Box218, Hawthorn, Vic 3122, Australia}

\begin{abstract}
Theoretical integrated broad-band colors ranging from far-UV to near-IR
have been computed for old stellar systems from our evolutionary population 
synthesis code. 
These models take into account, for the first time, the detailed systematic 
variation of horizontal-branch (HB) morphology with age and metallicity. 
Our models show that some temperature-sensitive color indices are 
significantly affected by the presence of blue HB stars. 
In particular, $B$ $-$ $V$ does not become monotonically 
redder as metallicity increases at given ages, but becomes bluer by 
as much as $\sim$ 0.15 mag because of the contribution from blue HB stars. 
Similar trends are also found in the Washington photometric system. 
In addition to appropriate age-sensitive spectrophotometric 
indices, the use of far-UV to optical colors is proposed as a powerful 
age diagnostic for old stellar systems with differing HB morphologies.

Our models are calibrated in the $B$ $-$ $V$, 
$V$ $-$ $I$, $C$ $-$ $T_{1}$, and $M$ $-$ $T_{1}$ vs. [Fe/H] planes, 
using low-reddened Galactic globular clusters (GCs) 
[$E$($B$ $-$ $V$) $<$ 0.2] and the relative age difference between 
the older inner halo Galactic GCs 
and younger outer halo counterparts is well reproduced. 
Several empirical linear color-metallicity transformation relations 
are assessed with our models and it is noted that they may not be 
safely used to estimate metallicity if there are sizable age differences 
amongst GCs within and between galaxies. M31 GCs are found to be 
fundamentally similar to those in the Milky Way, not only in the 
optical to near-IR range, but also in the UV range. For globular cluster 
systems in two nearby giant ellipticals, M87 and NGC 1399, the current 
available photometric data in the literature does not appear 
sufficient to provide robust age discrimination. 
It is anticipated, however, that the detailed population models 
presented here coupled with further precise spectrophotometric 
observations of globular cluster systems in external galaxies 
from the large ground-based telescopes and space UV facilities will 
enable us to accurately estimate their ages and metallicities.
\end{abstract}


\section{INTRODUCTION}

When stellar systems can be resolved into individual stars such as 
open and globular star clusters in the Milky Way and Local Group galaxies, 
one can derive their ages and metallicities via isochrone fitting to 
the main sequence turnoff and red giant branch in the color-magnitude diagram 
and fitting luminosity functions to the white dwarf cooling sequence 
(Hansen et al. 2002). 
This luxury does not exist for more distant systems, however, so 
one must rely upon their integrated colors or spectra. Calibrating  
theoretical spectrophotometric quantities using local, resolved systems is 
then surely a necessary step (e.g., Gibson et al. 1999).

Since local and nearby extragalactic 
globular clusters (GCs) are easily detected and feasible 
for analysis due to their rather simple nature, there have been many 
photometric studies of globular cluster systems in external galaxies 
which have revealed important information regarding the link between GC systems 
and their host galaxies (see Ashman \& Zepf 1998 for a recent review). 
Information regarding their relative ages within a given galaxy or between 
galaxies is particularly relevant for studying the hierarchical formation 
of galaxies. 
As globular cluster ages and metallicities become better constrained, 
the opportunities for understanding the process of galaxy formation 
and its subsequent evolution can only increase.

In recent years, there have been many efforts to develop evolutionary 
population synthesis models (e.g., Bruzual \& Charlot 1993; Worthey 1994; 
Buzzoni 1995; Vazdekis et al. 1996; Maraston 1998; Lee 2001a, 2001b) in order 
to analyze the integrated spectrophotometric quantities from globular clusters 
and galaxies and to derive their mean ages and metallicities. 
It is true that those integrated quantities are indeed determined by 
age and metallicity of those stellar systems, but it also should be kept in 
mind that they are the two most arguably dominant parameters that determine 
globular clusters' horizontal-branch (HB) morphologies. Therefore, 
the integrated colors of GCs should be affected to some degree 
by differing HB morphologies at the similar metallicity 
(the `second parameter' effect). 
It is now generally accepted that age is the {\it global} second parameter 
that controls HB morphology [after the first parameter, metallicity 
-- Lee, Demarque, \& Zinn 1994; Sarajedini, Chaboyer, \& Demarque 1997; 
Salaris \& Weiss 2002], although for some GCs a third (or more) parameter may 
be needed to explain their peculiar HB morphologies [such as 
blue tail phenomenon -- Recio-Blanco et al. 2002]. 

We are extending this second parameter analysis to broader age ranges 
in this work, applicable even to potentially older stellar systems 
than we have in the Milky Way, and investigate how variation of 
HB morphology influences the integrated photometric quantities. 
With a better understanding of post-main sequence stellar evolution, 
it is now feasible to study the systematic effects of those 
evolved stars on the integrated spectrophotometric values. One vivid example 
has already appeared in Lee, Yoon, \& Lee (2000; hereafter Paper I) 
for the realistic assessment of ages of old stellar systems 
using the H$\beta$ index. 
A unique aspect of these models lies in its treatment of the systematic 
variation of HB morphology as a function of age and metallicity. 
In this paper, we present a detailed quantitative analysis toward the effects 
of HB morphology on theoretical integrated broad-band colors ranging from 
far-UV to near-IR at various stages of age and metallicity. 
In addition, integrated broad-band colors that have been used 
to estimate the metallicity of globular clusters are examined with 
our models in order to check the veracity of the so-called color-metallicity 
transformation relations that are widely used, but based solely upon Galactic 
GCs. 

In Section 2, our population models developed with and without 
consideration of HB stars are presented and 
compared with those of other groups. We first consider optical to near-IR 
colors, including those using the Washington photometric system, 
an investigation of UV to optical colors is then presented. 
Section 3 corroborates the validity of 
our models with HB stars by calibrating them using a sample of 
relatively low-reddened Galactic GCs and 
our results are then used to assess several color-metallicity 
transformation relations. We compare our models 
with available UV to near-IR data for GC systems 
in M31, M87, and NGC 1399 (e.g., Cohen, Blakeslee, \& Ryzhov 1998; 
Kissler-Patig et al. 1998; Forbes et al. 2001). 
In most cases, we have specially sought age-sensitive colors, 
to combine with spectroscopic metallicity information. 
A discussion of the implications of our work is provided in Section 4.

\section{POPULATION MODELS OF INTEGRATED BROADBAND COLORS
         WITH AND WITHOUT HORIZONTAL-BRANCH STARS}

The present models were constructed, in essence, from the same 
evolutionary population synthesis code, developed to study 
the stellar populations of globular clusters (Paper I). 
In summary, we have used the Yale Isochrones (Demarque et al. 1996) 
rescaled for $\alpha$-element enhancement\footnote{[$\alpha$/Fe] = 0.4 for 
[Fe/H] $<$ $-$1.3, [$\alpha$/Fe] goes from 0.4 to 0.0 as [Fe/H] goes 
from $-$1.3 to 0.0,
and [$\alpha$/Fe]=0.0 for [Fe/H] $>$ 0.0 (Wheeler, Sneden, \& Truran 1989).} 
(Salaris, Chieffi, \& Straniero 1993) 
and the HB evolutionary tracks by Yi, Demarque, \& Kim (1997). 
For metal-rich ([Fe/H] $>$ $-$0.5) populations, the value of the helium 
enrichment parameter, $\Delta${\it Y}/$\Delta${\it Z} = 2, 
was assumed. The standard Salpeter (1955) 
initial mass function was adopted for calculating the relative number 
of stars along the isochrones. For the conversion from theoretical 
quantities to observable quantities, we employed the 
stellar library of Lejeune, Cuisinier, \& Buser (1997, 1998)
in order to cover the largest possible range 
in stellar parameters such as metallicity, temperature, and gravity.
As discussed in Paper I, we have rescaled the isochrone ages, such that 
the 15 Gyr isochrones now corresponds to present day 12 Gyr old 
populations\footnote{According to Fig. 14 and Table 7 of Yi et al. (2001),
it is assured that our treatment of age reduction effect by 3 Gyr is basically
correct, though some conundrum remains as the trend may be reversed 
in the metal-rich regime.}
and $\Delta${\it t} = 0 corresponds to t $\simeq$ 12 Gyr. 

The treatment of the detailed systematic HB morphology variation 
with age and metallicity is also compatible with that of 
Paper I. Figure 1 presents our model isochrones 
as a function of HB morphology [(B$-$R)/(B+V+R), where 
B, V, and R are the numbers of blue HB stars, RR Lyrae variables, 
and red HB stars, respectively; Lee et al. 1994] and metallicity 
[Fe/H]. The Galactic GC data of Lee et al. (1994) are overplotted. 
For NGC 4590 and NGC 6656, the updated HB types are adopted 
from the Harris (1996, June 1999 version) compilation. 
In order to match the observed HB morphology of inner halo Galactic GCs 
(Galactocentric radius $\leq$ 8 kpc) at their currently favored mean age 
($\sim$ 12 Gyr, $\Delta${\it t} = 0 Gyr) in light of the distance 
scale suggested by {\it Hipparcos} (e.g., Chaboyer et al. 1998), 
the value of $\eta$, the parameter in Reimers (1975) law, was 
taken to be $\eta$ = 0.65. We then investigate the age range 
$-$ 4 Gyr $\leq$ $\Delta${\it t} $\leq$ + 4 Gyr. In Figure 1, 
the larger symbols are the relatively low-reddened Galactic GCs 
[$E$($B$ $-$ $V$) $<$ 0.2] that will be used to calibrate our models 
in Section 3.1. It is useful to note here 
that the inner halo Galactic GCs (filled circles) are not only
systematically older, but also more tightly grouped along our model 
\footnote{The filled symbol that lies significantly away from the inner
halo locus in Fig. 1 is NGC 6584. According to Table 5 of 
Dinescu, Girard, \& van 
Altena (1999), NGC 6584 is of very much eccentric orbital parameters
($R_{a}$ $\sim$ 13 kpc, $R_{p}$ $\sim$ 1 kpc) and therefore
it may not belong to the true inner halo cluster category (see also 
Section 4. of Lee et al. (1994) for more detailed description). Actually, 
taking NGC 6584 as an outer halo cluster could make our assertion stronger.}
isochrone than those in the outer halo (open circles). 
   
We have not taken into account the effects of extended blue tails on the HB, or 
blue stragglers, in this work, as it remains difficult to quantify them 
systematically with age and metallicity. We recognize though 
that some UV indices and colors may be affected by their inclusion. 
Furthermore, except for the case of UV to optical colors, 
post-asymptotic giant branch (PAGB) stars are not considered here.

\subsection{OPTICAL TO NEAR-IR RANGE COLORS}
                                
Here we address the theoretical integrated broad-band colors in the 
standard $UBVRIK$ filter system. The following is a brief summary of 
the computational procedure. First, for each star (or group of stars, 
after binning) with a given metallicity, temperature, and gravity, 
we find the corresponding colors and the bolometric correction values from 
the color tables of Lejeune et al. (1998). 
Second, these are converted into each bandpass' luminosity. Finally, 
after each bandpass' luminosity has been summed, for a simple 
stellar population of a given metallicity and age, we compute the 
integrated photometric magnitudes for each bandpass. Integrated 
broad-band colors are then calculated from these computed magnitudes.

Figure 2 presents our theoretical predictions for the integrated 
broad-band colors, 
(a) $U$ $-$ $V$, (b) $B$ $-$ $V$, (c) $V$ $-$ $I$, and (d) $V$ $-$ $K$ 
as a function of [Fe/H]. We find that 
our models with HB stars (thick lines) show quantifiable differences 
from those without HB stars (thin lines), notably 
in Figure 2b. {\it Because of the contribution from blue HB stars, 
$B$ $-$ $V$ does not become monotonically redder as metallicity 
increases at given ages, but becomes bluer by as much as $\sim$ 0.15 mag}. 
Similar ``wavy" features\footnote{Selected synthetic CMDs in Fig. 5 of 
Lee et al. (2000) would be helpful to understand this ``wavy" features.} 
have been seen in the temperature-sensitive H$\beta$ index (Paper I).
Other colors appear to be affected to some degree by our treatment
of HB morphology variations, but the effects are 
not as prominent in the age-range investigated here. 
In the succeeding investigation, 
we will mainly concentrate on $B$ $-$ $V$ and $V$ $-$ $I$ 
because these two colors are two of the most frequently 
referred to in the literature.

There have been related studies of the 
theoretical integrated broad-band 
colors in recent years (e.g., Worthey 1994; Kurth, Fritze-von 
Alvensleben, \& Fricke 1999; Brocato et al. 2000), however, 
the different stellar physics and calibrations result in 
inevitable differences in their predictions. 
In particular, the different treatment of the HB is noteworthy 
and may result in significant differences in the models.
For instance, Worthey (1994; W94) treated 
HB stars as a rather simple red clump, providing results 
similar to those of our models 
without HB stars (cf. http://astro.wsu.edu/worthey/). 
Both Kurth et al. (1999; KFF99) and Brocato et al. (2000; BCPR00)
include HB stars in their models 
but their values of $\eta$, the efficiency of mass loss, are different 
($\eta$ = 0.35 in KFF99 and $\eta$ = 0.4 in BCPR00) from ours 
($\eta$ = 0.65) probably because 
they adopted $\sim$ 15 Gyr for the absolute age of Galactic GCs 
(see Table 1 of Yi et al. 1999 for the different choice of $\eta$ 
in connection with that of the absolute age of Galactic GCs). 
For example, Figure 3 shows significantly different HB morphologies at the same 
metallicity and similar age as different values of $\eta$ (different 
amounts of mass loss) are adopted. 
The Padova isochrones (Girardi et al. 2000) 
employ the smaller value of $\eta$ = 0.4 compared 
to that which we use in this study 
(open circles and small points)\footnote{The solid line in Fig. 3 represents
the HB evolutionary track and the dashed line does that of post-HB.
For this study, we present the case with HB stars only. The study
with post-HB stars can be found in Lee (2001a).} 
and seemingly no mass dispersion is taken into account (triangles). 
Combination of these issues results in different model predictions, 
particularly for temperature-sensitive indices. 
BCPR00 tested several values of $\eta$ 
and showed that higher values of $\eta$ gave bluer colors (see their Fig. 8). 

Despite these differences, it is interesting to note that both 
KFF99 and BCPR00 do show slight crossovers among different ages 
at the low metallicity region in the $B$ $-$ $V$ vs. [Fe/H] plane 
(see Fig. 3 of KFF99 and Fig. 9 of BCPR00). 
In addition, it is noted that almost all models reveal some non-linear 
feature in the $V$ $-$ $I$ vs. [Fe/H] plane, being 
flatter in the low metallicity range, [Fe/H] $<$ $-$ 1.0. The importance 
of this non-linearity will be addressed further in Section 3.1
when the empirical {\it linear} color-metallicity relations are 
assessed with our models.

\subsection{WASHINGTON PHOTOMETRIC SYSTEM}

The Washington photometric system introduced by Canterna (1976) was 
originally developed as an optimum system for deriving 
temperatures ($T_{1}$ and $T_{2}$ filters) and abundances ($C$ and $M$ filters) 
for late-type giant stars. In essence, the $C$ filter covers about two thirds of 
the $U$ and $B$ filter, and the $M$ filter has its peak between the $B$ and $V$ 
filters and covers about half of each filter. 
The $T_{1}$ and $T_{2}$ filters are 
almost identical to the $R$ and $I$ filters, respectively (see, 
Fig. 1 of Lejeune \& Buser 1996). Because of its efficiency with the 
broader bandpasses than the commonly used $UBVRI$, the popularity of 
this system has grown in recent years, 
when obtaining integrated colors of GC systems 
in distant galaxies (e.g., Geisler \& Forte 1990; Lee \& Geisler 1993; 
Secker et al. 1995; Zepf et al. 1995; Ostrov et al. 1998). 
Perhaps the most important use of this system is in the determination of 
metallicities for distant GCs from their 
integrated colors, especially ($C$ $-$ $T_{1}$), with its very 
long color baseline (Geisler \& Forte 1990). 

Although the number of observational studies using the Washington 
photometry has been increasing, little has yet been done 
in modeling the colors for the system, 
at least in comparison with the extant body 
of model colors for the $UBVRI$ system. In this context, we investigate 
the theoretical integrated colors using the Washington photometric 
system. The computation of integrated colors 
using the Washington system required some modifications to the procedure 
described in Section 2.1. First, 
we generate the integrated spectral energy distributions (SEDs) 
for a simple stellar population of a given metallicity and age 
using the flux tables (Lejeune et al. 1998). 
Then the fluxes within each bandpass of 
the Washington photometric system are integrated using 
the relative filter response functions, and we calculate the integrated 
magnitudes and colors. We have taken 
the response functions of the Washington system from KPNO homepage 
(http://www.noao.edu). 

Figure 4 shows effects of HB stars on (a) $M$ $-$ $T_{1}$  
and (b) $C$ $-$ $T_{1}$ colors, as predicted from our models, 
as a function of [Fe/H]. Thin lines represent models 
without HB stars, while thick lines include the effects of HB stars, 
based on the model loci of Figure 1. It is found that both 
$M$ $-$ $T_{1}$ and $C$ $-$ $T_{1}$ colors are affected by blue HB stars and 
become bluer by $\sim$ 0.08 mag and by $\sim$ 0.2 mag, respectively, 
than the models without HB stars, similar 
to the feature seen in Figure 2b [($B$ $-$ $V$)] and the H$\beta$ index 
seen in Figure 6 of Paper I. Geisler, Lee, \& Kim (1996) 
examined the age, HB morphology, and metallicity 
sensitivities of the integrated $C$ $-$ $T_{1}$ color, based on 
the work of Cellone \& Forte (1996) and concluded it was 
an efficient metal abundance index, in contrast, we have found that 
the integrated $C$ $-$ $T_{1}$ is, to some degree, affected 
by the systematic variation of HB morphology with age and metallicity.
It is difficult to compare our results directly with those of 
Cellone \& Forte (1996; CF96), because CF96 
give their outputs only at two given ages (5 and 15 Gyr) 
and base their predictions upon the older Buzzoni (1989, 1995) 
population synthesis code. 
However, it is evident even in CF96 
that the predicted $C$ $-$ $T_{1}$ with an intermediate HB morphology 
is bluer than that of a red HB, at the same metallicity
(see Fig. 7 of Geisler et al. 1996).
Theoretical modeling of the Washington photometric system 
based on a large sample of individual stars would undoubtedly 
give more credibility to this particular aspect of stellar population 
synthesis work (e.g., Lejeune \& Buser 1996).

\subsection{UV TO OPTICAL RANGE COLORS}

The computational procedure for the relevant UV bandpass magnitudes 
is almost identical to that which we described 
in Section 2.2. For the case of 
UV to optical colors of an old stellar population, however, 
the contribution from PAGB stars is considered. 
A small number of PAGB stars 
can have a significant impact upon the integrated UV flux. The PAGB 
contribution is somewhat stochastic by nature, due to small number statistics 
being driven primarily by their short lifetimes. 
For instance, the evolutionary lifetime of a 0.565 M$_{\odot}$ PAGB star 
is only 40,000 years (Sch\"onberner 1983) 
compared to about 100 Myrs for HB stars. We have quantified the expected number 
ratio of PAGB to HB stars based on the ratio of their 
evolutionary lifetimes after more than 10 simulations 
and added the corresponding averaged 
fluxes using Sch\"onberner's (1983) evolutionary track 
for a 0.565 M$_{\odot}$ PAGB star.

Figure 5 shows several integrated SEDs from 
our models, with and without PAGB stars. Three 
ages ($\Delta${\it t} = + 4, 0, and $-$ 4 Gyr) are illustrated 
at the given metallicity ([Fe/H] = $-$ 1.17). All fluxes have been normalized 
to unity at 5550 {\AA} and then converted into magnitudes. 
One can then easily estimate magnitude differences 
relative to the $V$-band from Figure 5.
For the ensuing investigation, 15 $-$ $V$ 
and 25 $-$ $V$ are defined as the magnitude differences 
between the far-UV and the near-UV band centered at 1550 {\AA} 
and 2490 {\AA}, respectively, with 150 {\AA} bandwidths, 
as indicated by the grey columns in Figure 5, and the visual band.
The dashed lines in Figure 5 ($\Delta${\it t} = $-$ 4 Gyr) show that 
old stellar systems with only red HB stars 
would have minimal far-UV flux without a PAGB contribution. 
From the dotted lines in Figure 5 ($\Delta${\it t} = + 4 Gyr), 
we can see that once hot blue HB stars begin to become important 
for very old systems, the PAGB contribution becomes increasingly insignificant. 
Consequently, it is seen that the older populations with 
the bluer HB stars should be relatively UV bright, especially 
in the far-UV range. In the near-UV range, however, it is seen that 
the contribution from PAGB stars is relatively unimportant.

In Figure 6, 15 $-$ $V$ and 25 $-$ $V$ colors from our models 
including contributions from PAGB stars 
are plotted against [Fe/H] for five relative ages 
($\Delta${\it t} = $-$ 4, $-$ 2, 0, + 2, and + 4 Gyr, respectively). 
The 15 $-$ $V$ vs. 
[Fe/H] plot (Figure 6a) is reminiscent of Figure 1, which suggests that the 
far-UV to optical colors also describe the HB morphology, 
in a manner similar to that of the optical to near-IR colors. 
This leads us to believe that far-UV photometry, in combination with 
optical data, will provide a useful age discriminant for GC systems. 
In Figure 6b, however, it is found that 
25 $-$ $V$ from our models is relatively 
insensitive to the variation of HB morphology, compared to 
15 $-$ $V$. This is because in the near-UV range 
the contribution from main sequence turnoff stars is 
not insignificant. These models will be tested using the available 
UV datasets of Galactic and M31 GCs in Section 3.3.

\section{COMPARISON WITH OBSERVATIONS}
    
\subsection{CALIBRATION VIA GLOBULAR CLUSTER SYSTEM
            IN THE MILKY WAY}

Having discussed the theoretical aspects of generating integrated 
broad-band colors in Section 2, we now provide an empirical calibration 
for the models using the Milky Way's GC system. We adopt Harris' (1996) 
dataset for calibration purposes, but restrict ourselves to those 
clusters with line-of-sight reddenings $E$($B$ $-$ $V$) $<$ 0.2. 
The more highly-reddened clusters possess increasingly uncertain photometry. 
When needed, the extinction law of Cardelli, Clayton, \& Mathis (1989) 
is adopted.
   
Figure 7 contrasts this sample of Galactic GCs (where 
filled circles represent inner halo clusters, and open circles, 
outer halo clusters) 
with our models in the (a) ($B$ $-$ $V$)$_{o}$, (b) ($V$ $-$ $I$)$_{o}$, 
(c) ($M$ $-$ $T_{1}$)$_{o}$, and (d) ($C$ $-$ $T_{1}$)$_{o}$ vs. [Fe/H] planes.
The integrated $B$ $-$ $V$ and $V$ $-$ $I$ colors of 
Galactic GCs have essentially negligible uncertainties, due to the intrinsic 
brightness of the sample; the largest uncertainties can be traced to the 
uncertainty in foreground reddening, which corresponds to $<$ 0.02 mag 
for this sample. The limited sample of Galactic GCs with Washington photometry 
(Harris \& Canterna 1977) is shown along with our models in Figure 7c and 7d. 
As a matter of fact, this is a first ever try to separate the Milky Way
globulars into subpopulations and to address their relative age differences
using broadband colors. It appears that 
our models reproduce the differences between inner and outer halo clusters 
noted in Figure 1, in the sense that 
the inner halo clusters are not only more tightly grouped along 
the isochrone than the outer halo counterparts, but also relatively older
from the eye fit.

In Figure 7, we applied 0.03 mag, 0.01 mag, and 0.05 mag zero-point 
offsets to the models in $B$ $-$ $V$, $M$ $-$ $T_{1}$, and $C$ $-$ $T_{1}$, 
respectively, in order to minimize the residuals between the models and 
the data. These offsets may arise from uncertainties in the adopted 
theoretical stellar atmospheres or our rough age reduction treatment 
by 3 Gyr or the color-effective temperature calibration 
(e.g., Girardi 2001; Kurucz 2001; Yi et al. 2001; Westera et al. 2002). 
We apply the same offsets for the comparison against extragalactic systems 
in Section 3.2. The models of Figure 7 support the hypothesis that the
realistic HB morphology inclusion is truly important, and 
the fact that even the small relative age 
difference between inner and outer halo clusters is rather 
satisfactorily reproduced ascertains that we are making some 
progress here over the previous attempts   
(e.g., Fig. 43 of W94, Figs. 3, 4 of KFF99, Figs. 6, 9 of BCPR00, and 
Fig. 7 of Geisler et al. 1996).

Before our models are applied to extragalactic GC systems, several 
empirical {\it linear} color-metallicity transformation relations 
in the literature 
were assessed (see Figure 8). First, in the [Fe/H] vs. 
($B$ $-$ $V$)$_{o}$ plane (Fig. 8a), Couture, Harris, \& Allwright's (1990) 
relation (CHA90) is shown with our 
models, ($B$ $-$ $V$)$_{o}$ = 0.200[Fe/H] + 0.971. The CHA90 relation 
is based on Galactic GCs with $E$($B$ $-$ $V$) $\leq$ 0.4. Fig. 8b shows the 
[Fe/H] vs. ($V$ $-$ $I$)$_{o}$ plane, with CHA90's relation, 
[($V$ $-$ $I$)$_{o}$ = 0.198[Fe/H] + 1.207], Kissler-Patig et al.'s 
(1998) relation (KBSFGH98), [[Fe/H] = $-$ 4.50 + 3.27($V$ $-$ $I$)], 
Kundu \& Whitmore's (1998) 
relation (KW98), [[Fe/H] = $-$ 5.89 + 4.72($V$ $-$ $I$)], 
and Harris et al.'s (2000) relation (HKHHP00), 
[($V$ $-$ $I$)$_{o}$ = 0.17[Fe/H] + 1.15] contrasted with 
our models. These latter relations are also made 
using the relatively low-reddened Galactic GCs, except that of 
KBSFGH98, who made their relation using the NGC 1399 GCs 
(see Section 3.2). The [Fe/H] vs. ($C$ $-$ $T_{1}$)$_{o}$ plane 
is shown in Fig. 8c, with Geisler \& Forte's (1990) relation 
(GF90), [[Fe/H] = 2.35($C$ $-$ $T_{1}$)$_{o}$ $-$ 4.39] compared 
with our models. GF90's relation is based upon 
the dataset of Harris \& Canterna (1977), using the Galactic GCs 
in the range $-$ 2.25 $\leq$ [Fe/H] $\leq$ $-$ 0.25. 

Because these color-metallicity transformation relations rely upon 
Galactic GCs in a limited range of color (0.6 $<$ ($V$ $-$ $I$)$_{o}$ $<$ 1.1), 
caution must be employed when extrapolating their use 
beyond this range (e.g., Harris et al. 
1992; Harris, Harris, \& McLaughlin 1998). Our models in Figure 8 suggest 
that $B$ $-$ $V$ may not be a good metallicity indicator and even 
$V$ $-$ $I$ should be used with extra caution to derive metallicity 
because of the model non-linearity. Geisler \& 
Forte (1990) demonstrated that the $C$ $-$ $T_{1}$ index is 
highly metallicity-sensitive, with a total range of 0.84 mag 
for the Galactic GCs, more than twice that of $B$ $-$ $V$ and 
easily converted to [Fe/H] metallicities on the Zinn (1985) 
scale with a mean standard deviation of 0.19 dex (cf. Harris \& Harris 2002). 
However, Figure 8c suggests that even 
the color-metallicity transformation relation using $C$ $-$ $T_{1}$ 
is subject to large uncertainty in metallicity estimation, 
if there are sizable age differences amongst GCs within and between 
galaxies.

\subsection{COMPARISON WITH GLOBULAR CLUSTER SYSTEMS
            IN M31, M87, AND NGC 1399}

We have demonstrated in Section 3.1 that detailed modeling 
of HB morphology is important 
in stellar population synthesis by calibrating our models 
using the sample of Galactic GCs. We now compare data from the 
M31, M87, and NGC 1399 GC systems with our models 
in order to investigate whether they show 
any systematic differences, particularly in terms of age, 
compared to their Galactic counterparts. 
In Paper I, we suggested that the GC systems of M87 and NGC 1399 are 
a few billion years older than the Milky Way system, based upon 
an analysis of the [Fe/H] vs. H$\beta$ plane. We now examine this claim 
using integrated broad-band colors, instead of H$\beta$.

For the M31 system, we use the Barmby et al. (2000) compilation.
As in the case of the Galactic GC system, we restrict 
the M31 sample to those with $E$($B$ $-$ $V$) $<$ 0.2
and $\sigma$[Fe/H] $<$ 0.2. 
Figure 9 compares the M31 GCs (filled squares) with our models in the (a) 
($B$ $-$ $V$)$_{o}$ and (b) ($V$ $-$ $I$)$_{o}$ vs. [Fe/H] planes; 
Galactic GCs with $E$($B$ $-$ $V$) $<$ 0.2 (open circles) 
are also plotted. The M31 GCs are reasonably 
well reproduced by our models and we suggest that there 
are no significant differences between the Galactic and M31 systems.

For the case of the M87 GC system, there are 
at least two photometric studies carried out in the $UBVRI$ 
system (Strom et al. 1981; Couture et al. 1990). 
Both datasets, however, suffer to some degree from uncertain 
photometric accuracy. These data, after combining with 
Cohen et al.'s (1998) spectroscopic metallicity determinations, 
are compared with our models in Figure 10. We adopt a foreground 
reddening of $E$($B$ $-$ $V$) = 0.02 (Burstein \& Heiles 1982), 
corresponding to $E$($C$ $-$ $T_{1}$) = 0.04.
   
A comparison of the Strom et al. (1981; SFHSWS81) photometry 
for the M87 GC system with our models is shown in Fig. 10(a) 
($U$ $-$ $B$)$_{o}$, 10(b) ($U$ $-$ $R$)$_{o}$, and 10(c) ($B$ $-$ $R$)$_{o}$. 
Here the brighter sample of data ($B$ $<$ 21) is highlighted 
with large filled circles, while the fainter sample ($B$ $<$ 22) 
is shown with small open circles. 
Strom et al. (1981) suggest that uncertainties in their color 
zero-points amount to 0.1 - 0.2 mag. We found that zero-point
offsets of 0.08 mag in $U$ $-$ $B$, 0.11 mag in 
$U$ $-$ $R$, and 0.01 mag in $B$ $-$ $R$ were necessary to 
match our models to their photographic photometry. The models 
are consistent with the data, particularly for the brighter clusters 
with smaller observational uncertainties. Although difficult to extract robust
age information from this sample, Figure 10c nevertheless shows that 
higher quality data paricularly in the regime of [Fe/H] $<$ $-$ 1
may allow for a useful age discrimination in the future.

Another comparison of M87 GCs (using photometric data 
from Couture et al. 1990) with our models in the 
($B$ $-$ $V$)$_{o}$ and ($V$ $-$ $I$)$_{o}$ vs. [Fe/H] planes 
is presented in Figure 11. It appears that this sample of M87 GCs  
also exhibits systematic zero-point offsets, 
as noted by Couture et al. of the order of $\pm$ 0.04 mag in color. 
With only 9 clusters in common between 
Couture et al. (1990) and Cohen et al. (1998), it is difficult 
to obtain useful age information of this system mainly because of the residual 
observational uncertainties. Additional high-quality photometric data 
(at least $B$, $V$, and $I$) is urgently required, in order to examine 
any potential age differences between GC systems in M87 and the Milky Way. 

The Washington CCD photometry for the M87 GCs is compared with our models
in Figure 12. Thirty-two clusters have both 
photometry (Lee \& Geisler 1993; LG93) and spectroscopy (Cohen et al. 1998). 
We have used only 25 of these here, excluding 
the C field (photometric uncertainty) and 
W field (non-photometric night) data of LG93; they are plotted in 
Figure 12(a) and 12(b). The typical photometric error 
in $C$ $-$ $T_{1}$ for the cluster sample is 0.06 mag, 
but far smaller for the brighter GCs. 
The typical uncertainty in spectroscopic metallicity is $\sim$ 0.2 dex, and 
is not displayed here for clarity. 
The QSNR, defined by Cohen et al. (1998) in their spectroscopic analysis, 
correlates with photometric magnitude in the sense that 
brighter clusters are of higher QSNR. The large filled circles of 
Figure 12 have QSNR $>$ 50, and are the relatively bright clusters 
with relatively high accuracy in photometry 
as well as in spectroscopy.  
It is seen from Figure 12 that even though our models 
trace the data reasonably well, 
the low quality of the M87 GCs photometry prevents us from any precise
age estimation.

In Figure 13, we present a 
comparison of the NGC 1399 GCs (photometric data from 
Kissler-Patig et al. 1997, [Fe/H] from Kissler-Patig et al. 1998) 
with our models in the [Fe/H] vs. ($V$ $-$ $I$)$_{o}$ plane. 
There appears to be a systematic offset between the data and our models. 
Kissler-Patig et al.'s (1998) color-metallicity transformation relation 
(Figure 8b) differs substantially from other comparable relations, 
therefore it would be worthwhile to get more number of fine photometry
of metal-rich clusters to address this seeming discrepancy.

\subsection{UV PHOTOMETRY}

We can conclude from Section 3.2 that reliable 
age information for extragalactic GC systems 
is difficult to obtain with the currently available 
optical to near-IR data. As suggested in Section 2.3 though, far-UV photometry 
would appear to be more useful as an age discriminant. Here, we examine 
whether our theoretical UV to optical colors can be applied 
to the available UV photometry of globular clusters in the Milky Way and M31.

In Figure 14, data for the Galactic GCs from 
the {\it Orbiting Astronomical Observatory} ({\it OAO 2}, 
filled circles) and the 
{\it Astronomical Netherlands Satellite} ({\it ANS}, 
filled squares) and those for the M31 GCs 
from the {\it Ultraviolet Imaging Telescope} ({\it UIT}, open circles) 
are plotted with our models. Triangles represent the relatively
highly-reddened clusters with $E$($B$ $-$ $V$) $>$ 0.2.
For the Galactic system, there are 
7 far-UV and 13 near-UV data points from the {\it OAO 2}, and 17 
far-UV and 22 near-UV data points from the {\it ANS} 
(Dorman, O'Connell, \& Rood 1995). 
Bohlin et al. (1993) report 43 near-UV and 4 far-UV detections 
of M31 GCs using the {\it UIT}. From their Table 2 \& 3, however,
only 17 near-UV and zero far-UV photometry are used here
after the exclusion of uncertain identifications (indicated by a colon
following the Bo ID in their tables) and considering only those 
have both [Fe/H] and $E$($B$ $-$ $V$) from Barmby et al. (2000).

The predicted (15 $-$ $V$)$_{o}$ colors from our models as a function of [Fe/H] 
are compared with Galactic GCs in the left panel of Figure 14. 
The {\it OAO 2} photometry is shown in Figure 14a, and 
Figure 14b presents that from the {\it ANS}. The errors 
associated with the {\it OAO 2} data amount to $\sim$ 0.5 mag, 
and are omitted in Figure 14a. As we compare it to Figure 1, 
it is noted that the observations fall within the suggested age range of our  
model predictions. Moreover, the far-UV flux from  
the well-known second parameter pair, M3 and M13 (depicted as crosses), is 
explicable with a relative age difference of $\sim$ 1.5 Gyr between the pair. 
This is consistent with the recent age estimation for these clusters 
based on their optical CCD photometry (Rey et al. 2001).
We also note that 47 Tuc (Figure 14a) is well reproduced 
by our models, verifying our estimation of the PAGB star 
contribution to the integrated far-UV flux.  
The use of far-UV to optical colors is, therefore, suggested as 
the most promising chronometer for GC systems, with the admitted caveat that 
there are a few peculiar GCs, such as NGC 6388 
and NGC 6441 with unexpected extended blue HB stars (Rich et al. 1997). 
If there are significant age differences between Milky Way and giant elliptical 
GC systems, in the sense of the latter being 2 - 3 Gyr older (as 
suggested in Paper I), 
then we predict that the majority of metal-rich clusters 
would be far-UV bright compared to Galactic counterparts. 
In this respect, the results from the recent HST 
far-UV photometry of GC systems in Virgo ellipticals 
(GO 8643 \& GO 8725) are highly anticipated. 

In the right panel of Figure 14, the (25 $-$ $V$)$_{o}$ colors of M31 GCs 
({\it UIT} data, open circles) are shown with our models 
as a function of [Fe/H]. In addition, the relevant data for Galactic GCs 
({\it OAO 2}; Figure 14c, {\it ANS}; Figure 14d) are overplotted 
for comparison. We have recalibrated the (25 $-$ $V$)$_{o}$ colors of the 
M31 GCs by adopting $V$ and $E$($B$ $-$ $V$) from Barmby et al. (2000).
Our models represent the near-UV photometry of M31 GCs, 
as well as those in the Milky Way, with no significant 
differences seen between the two systems, especially 
when disregarding the highly-reddened clusters (triangles). 
This again suggests that the M31 and Milky Way GC systems are fundamentally 
similar, in agreement with the analysis of optical to near-IR colors 
seen in Section 3.2, and consistent with the earlier conclusions of 
Bohlin et al. (1993). As discussed in Section 2.3, it is somewhat 
difficult to discriminate age differences of GC systems using the near-UV 
photometry alone. Far-UV photometry of the 
M31 GCs would be highly desirable 
to better quantify our findings. In this respect, a large 
UV photometry dataset for the M31 GC system 
from the upcoming {\it GALEX} (http://www.srl.caltech.edu/galex/) mission 
will become an invaluable resource.

\section{CONCLUSIONS \& DISCUSSION}

The primary goal of our paper was to investigate the effects of 
horizontal-branch stars on the integrated broad-band colors 
of old, simple, stellar populations. To do so, we have employed a 
unique, self-consistent, treatment of HB morphology as a function of 
age and metallicity. We have found that some temperature-sensitive 
integrated broad-band colors are significantly affected by the presence 
of blue HB stars within our investigated age range 
of $-$ 4 Gyr $\leq$ $\Delta${\it t} $\leq$ + 4 Gyr (i.e., 8 Gyr $\leq$ t 
$\leq$ 16 Gyr). The close agreement between our models 
and the relatively low-reddened inner halo Galactic GCs 
in both the Johnson-Cousins and Washington filter systems is 
encouraging. The use of far-UV to optical colors 
is suggested to be a powerful age-dating regime for 
old stellar systems, when coupled with realistic HB morphologies 
(Lee 2001a, 2001b). Future observations of GC systems in external galaxies 
from large ground-based telescopes and space UV facilities will 
enable us to quantify any systematic age differences between the 
various systems. A more sophisticated understanding of the theory governing 
mass loss and helium enrichment will also contribute to a better 
understanding of stellar age determination 
using far-UV photometry (O'Connell 1999).

In the case of the M31 GC system, our work suggests that it is not 
fundamentally different from the Galactic system, from the UV through to the 
near-IR. This is in contrast to several studies which have claimed 
that the metal-rich M31 GCs could be much younger than the 
metal-poor sample (e.g., Burstein et al. 1984; Barmby \& Huchra 2000). 
Future far-UV datasets to be provided, for example, by the {\it GALEX} mission 
should aid in resolving this controversy. Further Washington system photometry 
will also be valuable (Lee et al. 2001).

The study of extragalactic GC systems today is being driven primarily 
by an attempt at understanding the bimodal 
color distributions seen from many early-type
galaxies (e.g., Larsen et al. 2001; Kundu \& Whitmore 2001) 
as well as from some spirals (Forbes, Brodie, \& Larsen 2001). 
The origin of these blue and red subpopulations 
and the implication for the formation of their host galaxies remains unclear. 
If a relative age difference exists between the subpopulations, 
an important piece of the galaxy formation puzzle will have been found. 
Identifying useful age discriminants remains an important component of 
cosmology and galaxy formation. The broad-band photometric predictions 
described here, in combination with the spectroscopic discriminants 
of Paper I, should provide observers with the necessary tools to 
discriminate age differences amongst GC subpopulations. 
Finally, we conclude that several of the canonical 
{\it linear} color-metallicity transformation 
relations should be used with caution if there are sizable age differences 
amongst globular clusters within and between galaxies. 
It remains to be seen if bimodal color distributions can be interpreted 
as simple metallicity or age differences, or whether a more complicated 
interplay between metallicity and age must be involved.

\acknowledgments

It is a pleasure to thank P. Barmby for M31 GCs data, 
M. G. Lee for M87 GCs data, T. Lejeune for his updated 
color table, and S. K. Yi for many helpful discussions. 
We are also grateful to the anonymous referee for her/his detailed report 
that helped us improve this paper. Support for this work was  
provided by the Creative Research Initiatives 
Program of the Korean Ministry of Science and Technology. 
This work was also supported by the Post-doctoral Fellowship Program
of Korea Science \& Engineering Foundation (KOSEF). BKG acknowledges 
the support of the Australian Research Council through its Large Research 
Grant Program (A00105171).

\clearpage

\begin{figure}
\epsscale{.9}
\plotone{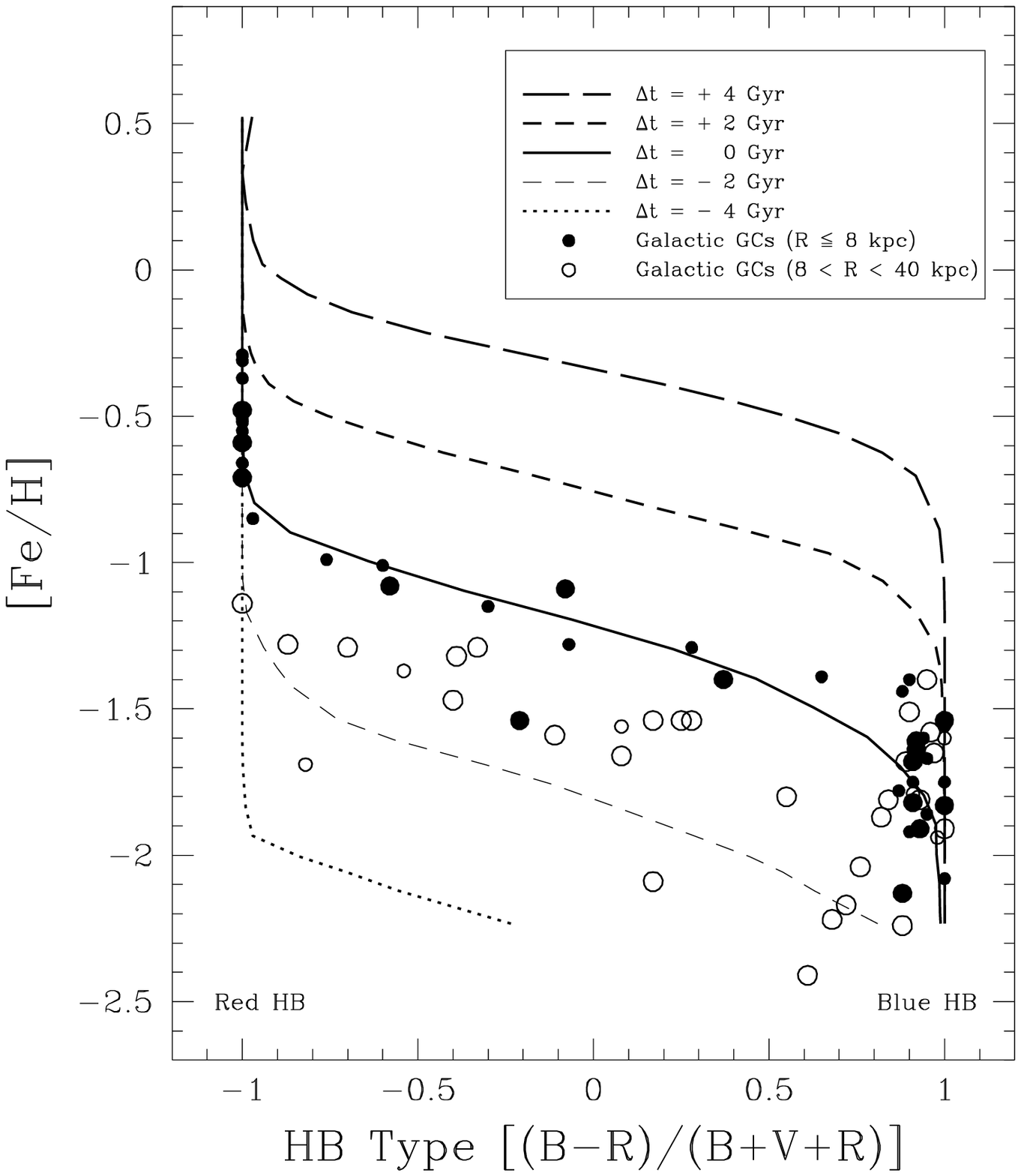}
\caption{The detailed systematic HB morphology variation was established 
by matching the tight correlations between the HB morphology type 
and [Fe/H] of the inner halo Galactic globular clusters 
(Galactocentric distance $\leq$ 8 kpc; filled symbols) 
at their currently favored mean age ($\sim$ 12 Gyr, $\Delta${\it t} = 0 Gyr; 
solid line). Open symbols represent the outer halo 
Galactic GCs. The dotted line and the 
short-dashed line (thinner one) represent the predicted model 
relationship for populations that are 4 Gyr and 2 Gyr younger, 
and the short-dashed (thicker one) and long-dashed lines 
represent that for populations that are 2 Gyr and 4 Gyr older 
than the inner halo Galactic GCs, respectively. 
The larger symbols are the relatively low-reddened Galactic GCs 
[$E$($B$ $-$ $V$) $<$ 0.2] that are plotted in Fig. 7.
Note that the inner halo Galactic GCs are not only systematically 
older, but also more tightly grouped along our model 
isochrone than those in the outer halo in this diagram (see text).
Data are from Lee, Demarque, \& Zinn (1994).}
\end{figure}

\begin{figure}
\plotone{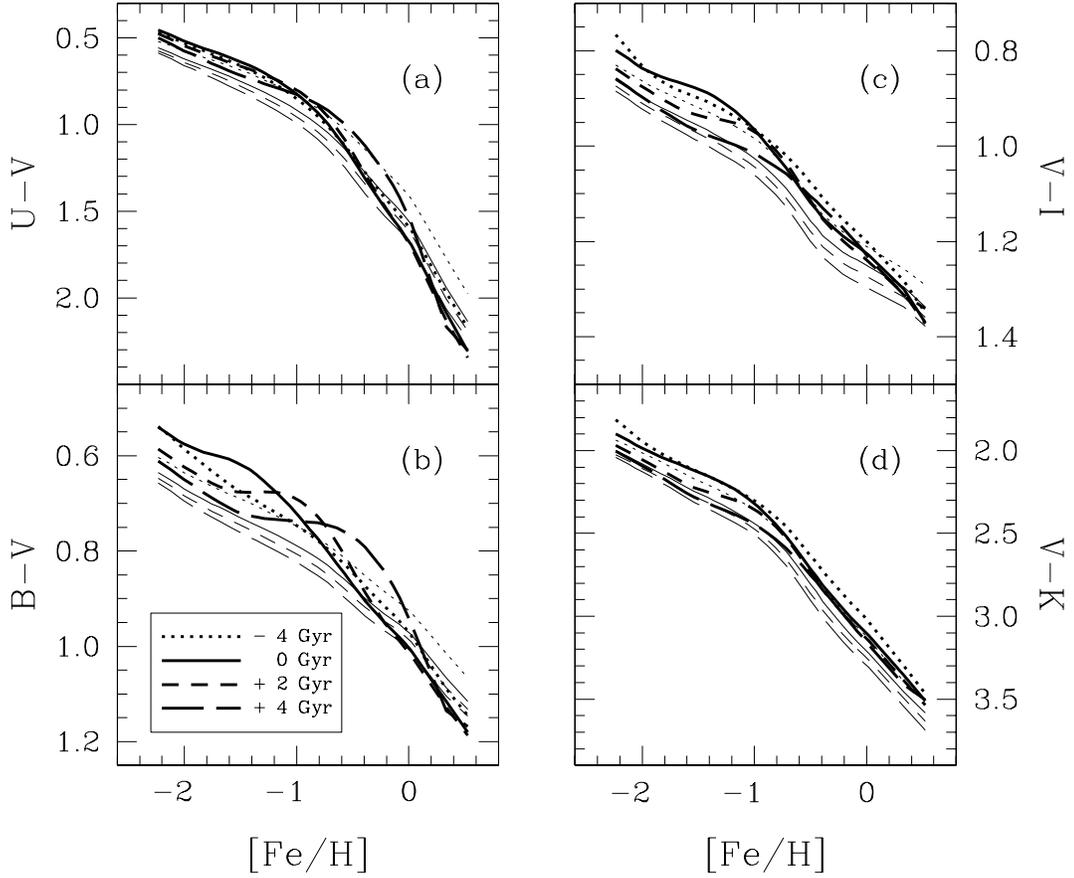}
\caption{Effects of HB stars on (a) $U$ $-$ $V$, (b) $B$ $-$ $V$, 
(c) $V$ $-$ $I$, 
and (d) $V$ $-$ $K$ colors as predicted from our models (see text). 
The colors are plotted against [Fe/H] for four relative ages 
($\Delta${\it t} = $-$ 4, 0, + 2, and + 4 Gyr, respectively). Thin lines 
represent models without HB stars, while thick lines represent 
those with HB stars based on the model loci of Fig. 1. 
Note that $B$ $-$ $V$ colors are particularly affected by blue HB stars and 
become bluer by $\sim$ 0.15 mag than the models without HB stars 
similar to the feature seen in the H$\beta$ index (Paper I).}
\end{figure}

\begin{figure}
\plotone{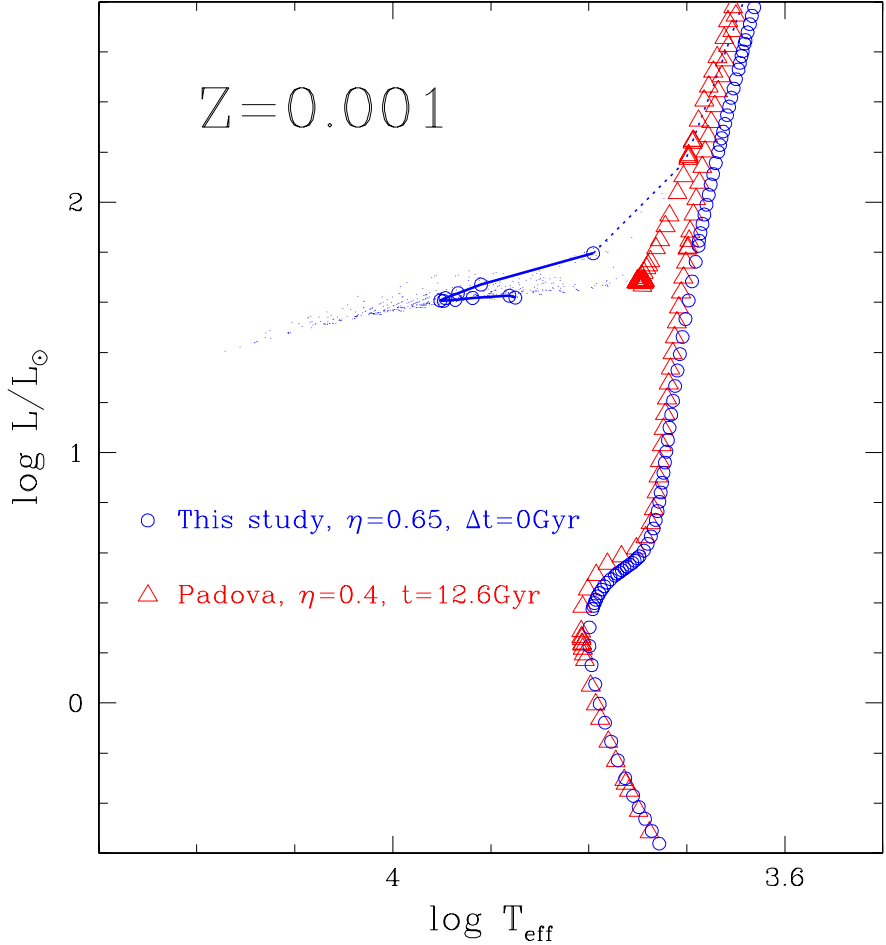}
\caption{A selected HR diagram that shows 
differences of this study's HB morphology (open circles)
from the Padova isochrone's one (triangles) 
at the same metal abundance (Z=0.001) and similar age. 
This takes place mainly because of adopting different amounts of 
mass loss (i.e., different values of $\eta$) and due to our treatment 
of mass dispersion (small points represent the realistic HB morphology
with Gaussian mass dispersion for our case). Combination of these issues
results in different model predictions,
particularly for temperature-sensitive indices. The evolutionary track
of our models' mean mass HB star is connected by lines (see text).}
\end{figure}

\begin{figure}
\epsscale{.7}
\plotone{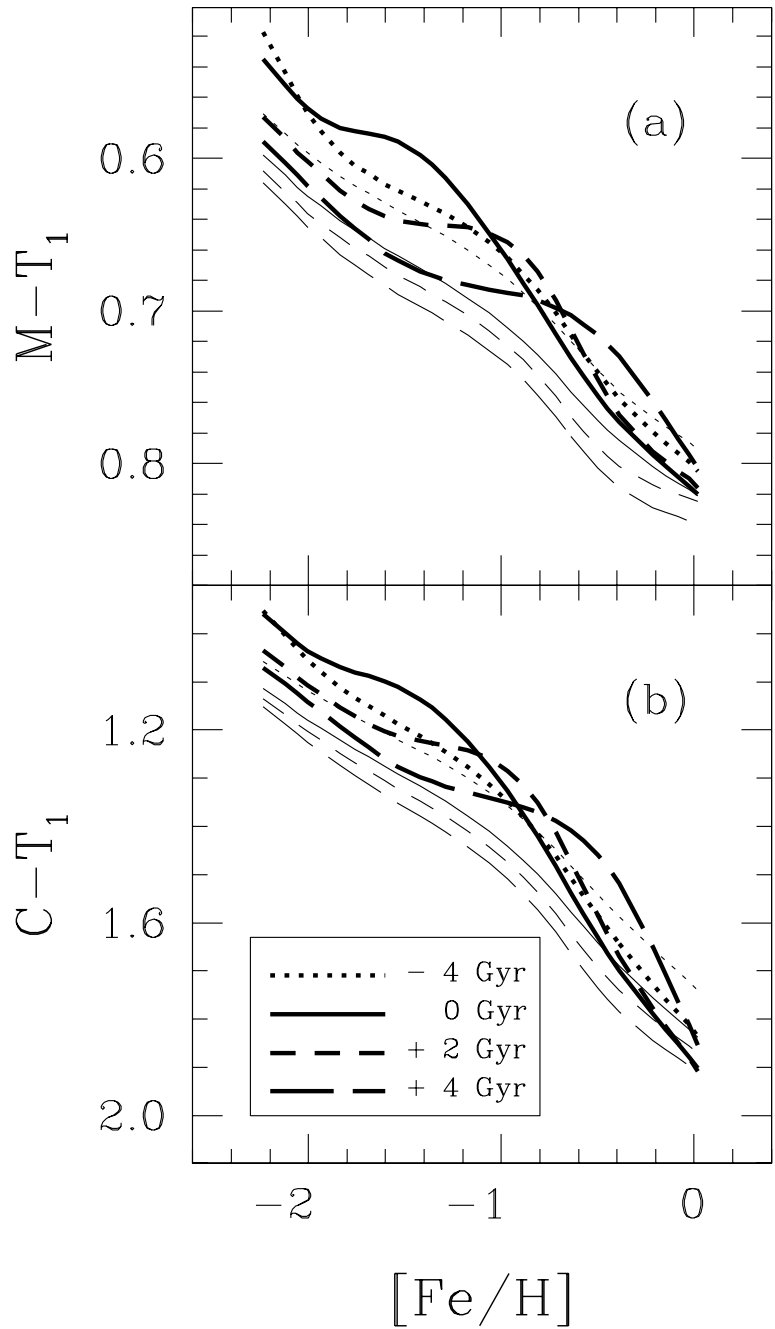}
\caption{Effects of HB stars on (a) $M$ $-$ $T_{1}$ and 
(b) $C$ $-$ $T_{1}$ colors 
as predicted from our models. The colors are plotted against [Fe/H] 
for four relative ages. Symbols are same as in Fig. 2.
Note that both $M$ $-$ $T_{1}$ and $C$ $-$ $T_{1}$ colors 
are affected by blue HB stars and become 
bluer by $\sim$ 0.08 mag and by $\sim$ 0.2 mag, respectively, 
than the models without 
HB stars similar to the feature seen in Fig. 2b.}
\end{figure}

\begin{figure}
\plotone{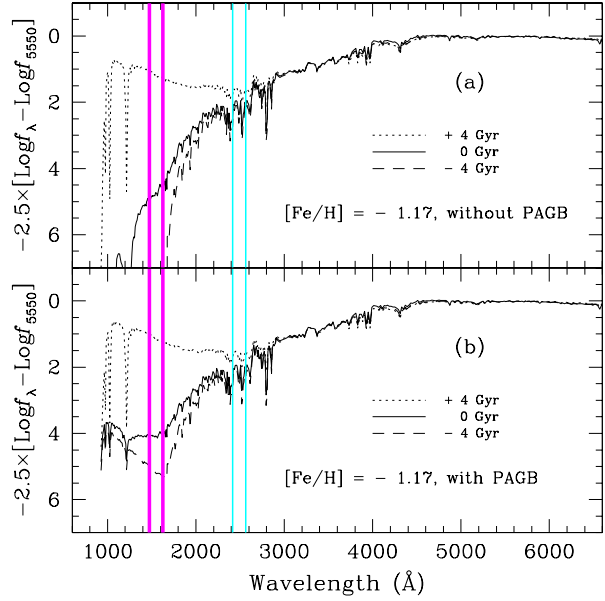}
\caption{The selected integrated SEDs from our models (a) without  
and (b) with including contribution from PAGB 
stars are drawn for three different ages ($\Delta${\it t} = + 4, 0, and 
$-$ 4 Gyr, respectively) at a given metallicity ([Fe/H] = $-$ 1.17).
The far-UV (thick grey column) and the near-UV (thin grey column)
bandpasses that we have employed in this study are indicated.
Note that once the old populations 
are fairly UV bright, especially in the far-UV range 
due to hot blue HB stars (dotted lines), the PAGB contribution 
becomes unimportant.}
\end{figure}

\begin{figure}
\plotone{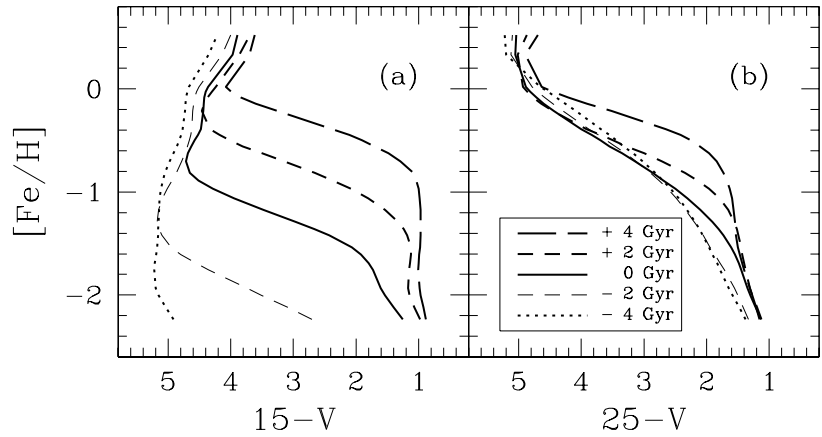}
\caption{(a) 15 $-$ $V$ and (b) 25 $-$ $V$ colors from our models 
including contributions from PAGB stars are plotted as a function of [Fe/H] 
for five relative ages. Symbols are same as in Fig. 1.
Note that 15 $-$ $V$ vs. [Fe/H] plot is reminiscent 
of the HB type vs. [Fe/H] plot drawn in Fig. 1.}
\end{figure}

\begin{figure}
\plotone{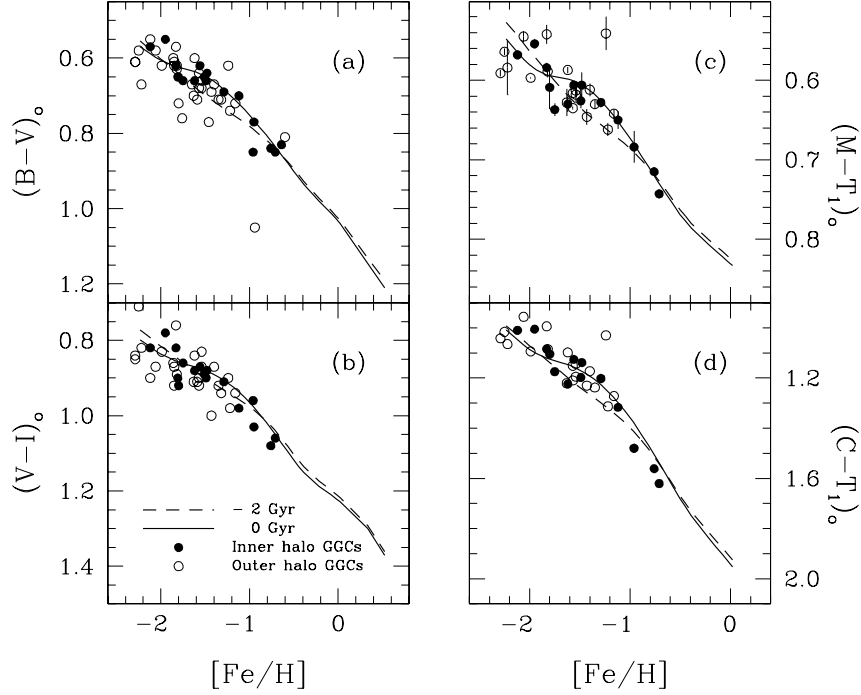}
\caption{The relatively low-reddened Galactic GCs 
[$E$($B$ $-$ $V$) $<$ 0.2, filled circles; inner halo clusters, 
open circles; outer halo clusters] are used to calibrate our models 
in the (a) ($B$ $-$ $V$)$_{o}$, (b) ($V$ $-$ $I$)$_{o}$, 
(c) ($M$ $-$ $T_{1}$)$_{o}$, 
and (d) ($C$ $-$ $T_{1}$)$_{o}$ vs. [Fe/H] planes.
Note that our models not only represent the observational data 
reasonably well but also 
reproduce the different trends, particularly in terms of the relative ages, 
between inner and outer halo clusters noted in Fig. 1. 
In $B$ $-$ $V$, $M$ $-$ $T_{1}$, and $C$ $-$ $T_{1}$, 0.03 mag,
0.01 mag, and 0.05 mag zero-point offsets were applied, respectively 
(see text).}
\end{figure}

\begin{figure}
\epsscale{.5}
\plotone{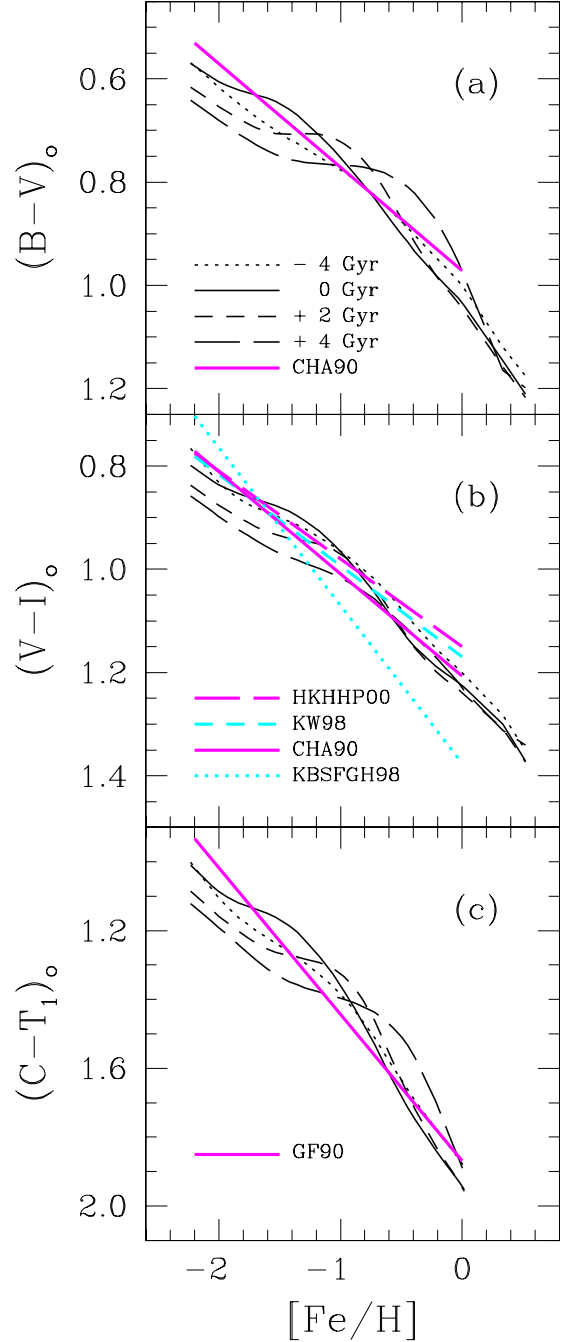}
\caption{Our models are contrasted with several {\it linear} color-metallicity 
transformation relations. Symbols 
for our models are shown in the panel (a). 
In the [Fe/H] vs. ($B$ $-$ $V$)$_{o}$ plane (a), 
Couture et al.'s (1990) relation (CHA90, grey solid line), 
in the [Fe/H] vs. ($V$ $-$ $I$)$_{o}$ 
plane (b), Couture et al.'s (1990) relation (CHA90, 
grey solid line), Kissler-Patig et al.'s (1998) relation 
(KBSFGH98, grey dotted line), Kundu \& Whitmore's (1998) relation 
(KW98, grey short-dashed line), and Harris et al.'s (2000) relation 
(HKHHP00, grey long-dashed line), and in the [Fe/H] vs. 
($C$ $-$ $T_{1}$)$_{o}$ plane (c), Geisler \& Forte's (1990) relation 
(GF90, grey solid line) are compared, respectively, with our 
models (see text).}
\end{figure}

\begin{figure}
\epsscale{.7}
\plotone{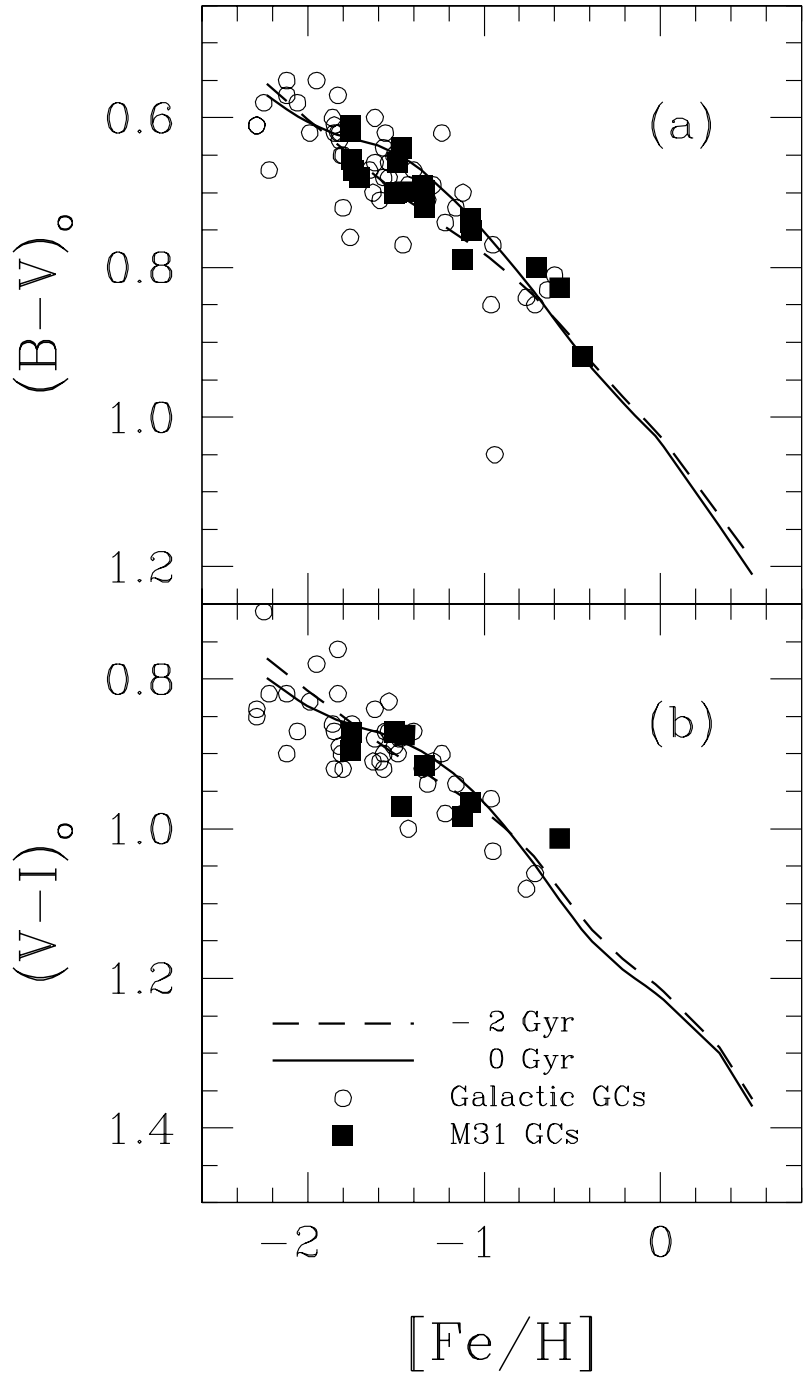}
\caption{The M31 GCs (filled squares; $E$($B$ $-$ $V$) $<$ 0.2, 
$\sigma$[Fe/H] $<$ 0.2, 
data from Barmby et al. 2000) are compared with our models in the (a) 
($B$ $-$ $V$)$_{o}$ and (b) ($V$ $-$ $I$)$_{o}$ vs. [Fe/H] planes. 
For comparison, Galactic GCs (open circles; 
$E$($B$ $-$ $V$) $<$ 0.2) are also plotted. 
Note that our models reproduce the observational data reasonably well 
and there are no significant differences between the Galactic and M31 systems.}
\end{figure}

\begin{figure}
\epsscale{.5}
\plotone{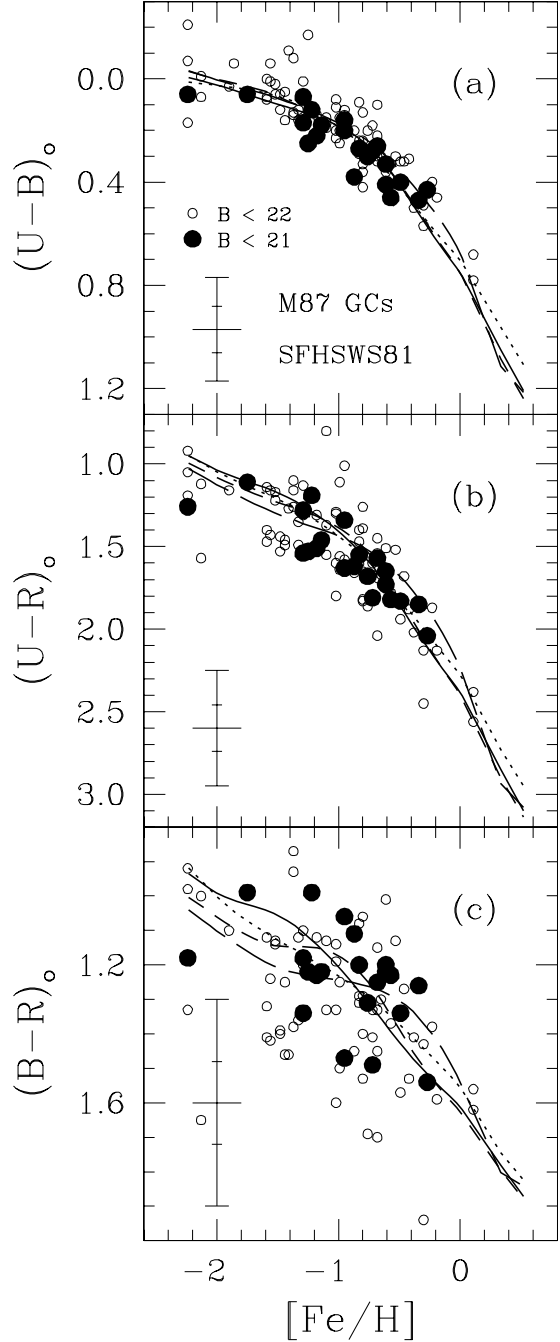}
\caption{The M87 GCs (small open circles; $B$ $<$ 22, large filled circles; 
$B$ $<$ 21, photometric data from Strom et al. 1981 (SFHSWS81), [Fe/H] from 
Cohen et al. 1998) are compared with our models in the (a) 
($U$ $-$ $B$)$_{o}$, (b) ($U$ $-$ $R$)$_{o}$, and (c) 
($B$ $-$ $R$)$_{o}$ vs. [Fe/H] planes. 
Symbols for our models are same as in Fig. 2. Photometric errors 
are displayed for clusters with $B$ $<$ 21 (small error bars) and with 
$B$ $<$ 22 (large error bars), respectively. Note that our models are 
consistent with the data, though it is difficult to extract robust 
age information from this sample because of 
the still large observational errors (see text).}
\end{figure}

\begin{figure}
\epsscale{.7}
\plotone{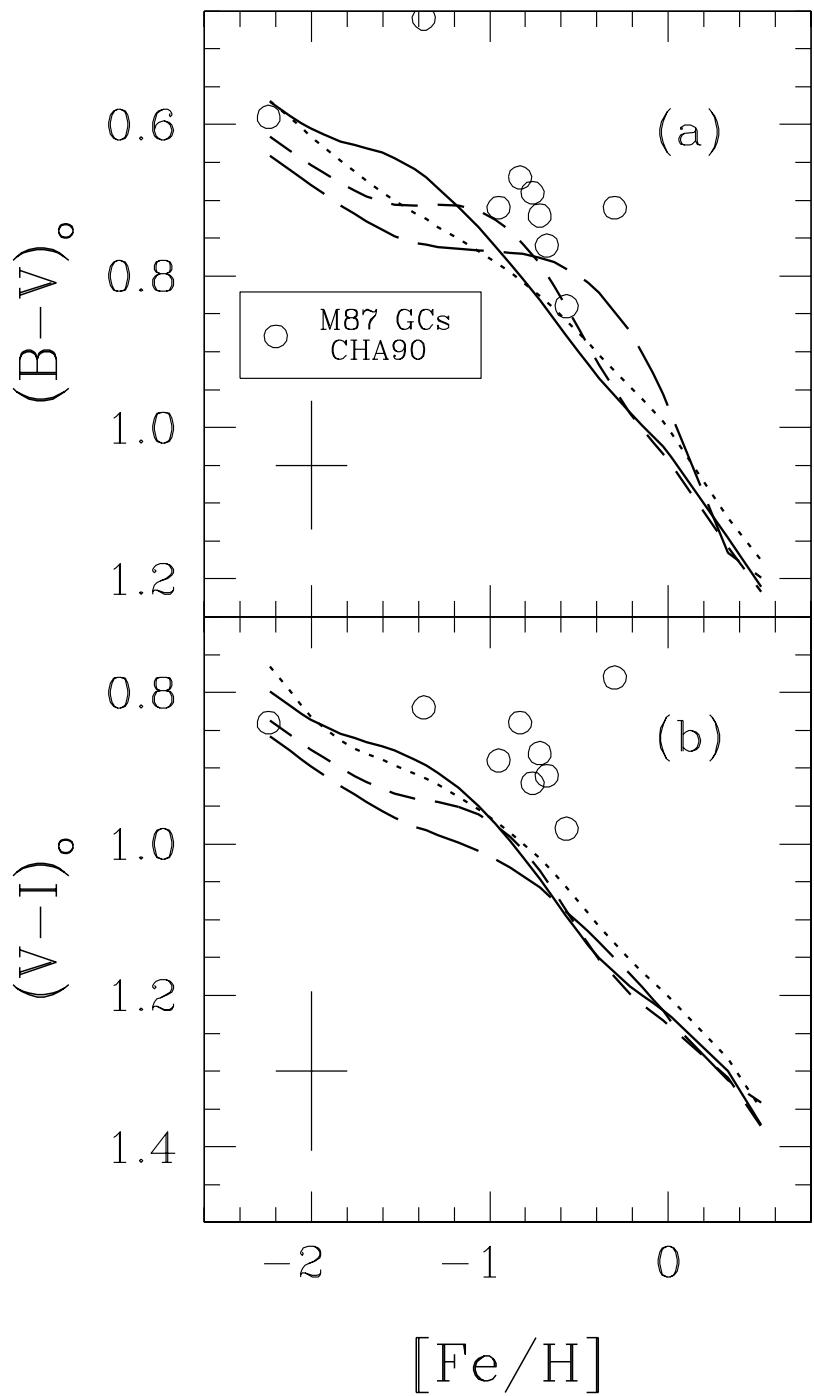}
\caption{The M87 GCs (photometric data from 
Couture et al. 1990 (CHA90), [Fe/H] from 
Cohen et al. 1998) are compared with our models in the (a) 
($B$ $-$ $V$)$_{o}$ and (b) ($V$ $-$ $I$)$_{o}$ vs. [Fe/H] planes. 
Symbols for our models are same as in Fig. 2. 
Note that there seem to be systematic zero-point offsets (see text).}
\end{figure}

\begin{figure}
\epsscale{.7}
\plotone{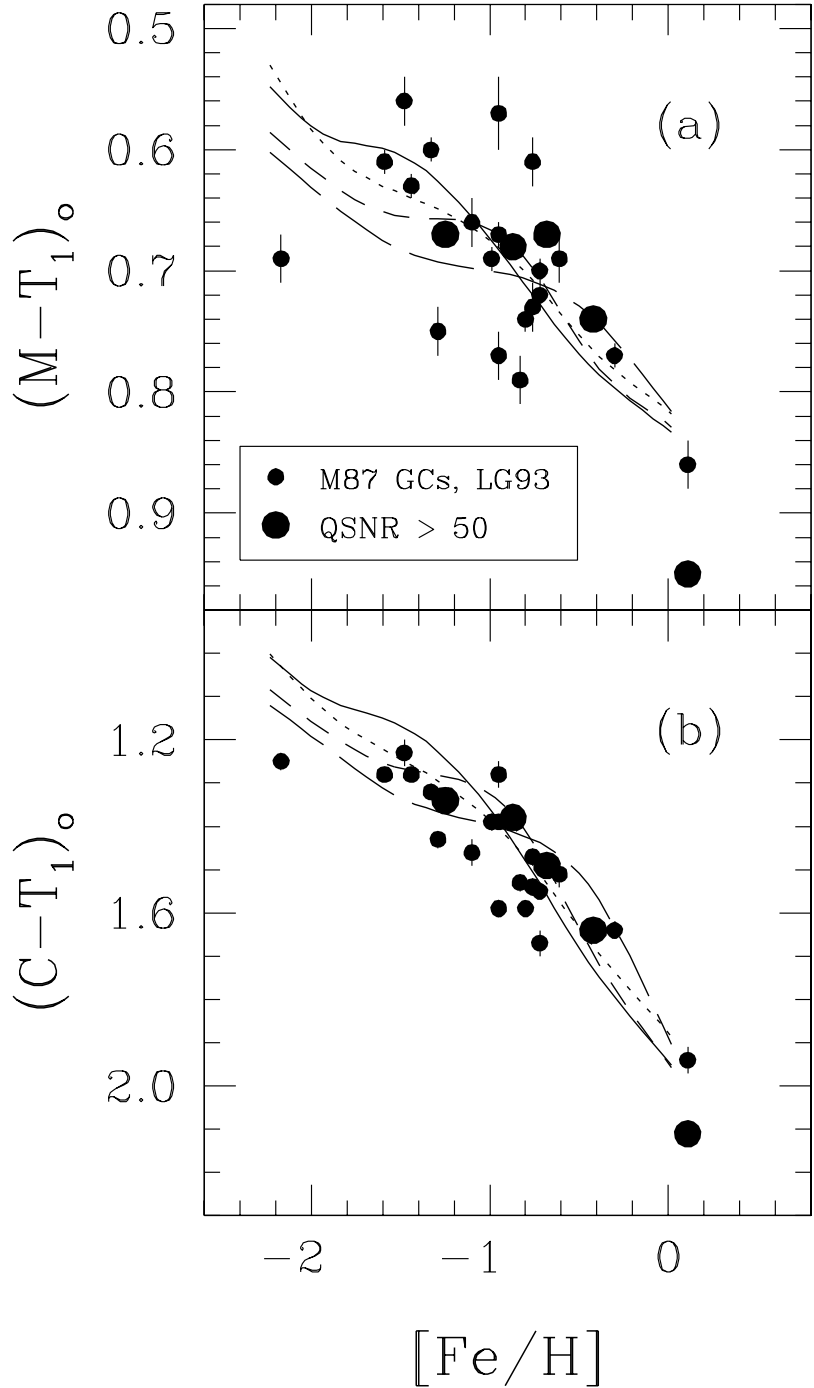}
\caption{The M87 GCs (photometric data from Lee \& Geisler 1993 (LG93), 
[Fe/H] from Cohen et al. 1998 (large symbols, QSNR $>$ 50)) are compared 
with our models in the (a) 
($M$ $-$ $T_{1}$)$_{o}$ and (b) ($C$ $-$ $T_{1}$)$_{o}$ vs. [Fe/H] planes. 
Symbols for our models are same as in Fig. 2. 
Note that our models trace the data reasonably well, although 
the low quality of the M87 GCs photometry prevents us from any precise
age estimation.}
\end{figure}

\begin{figure}
\plotone{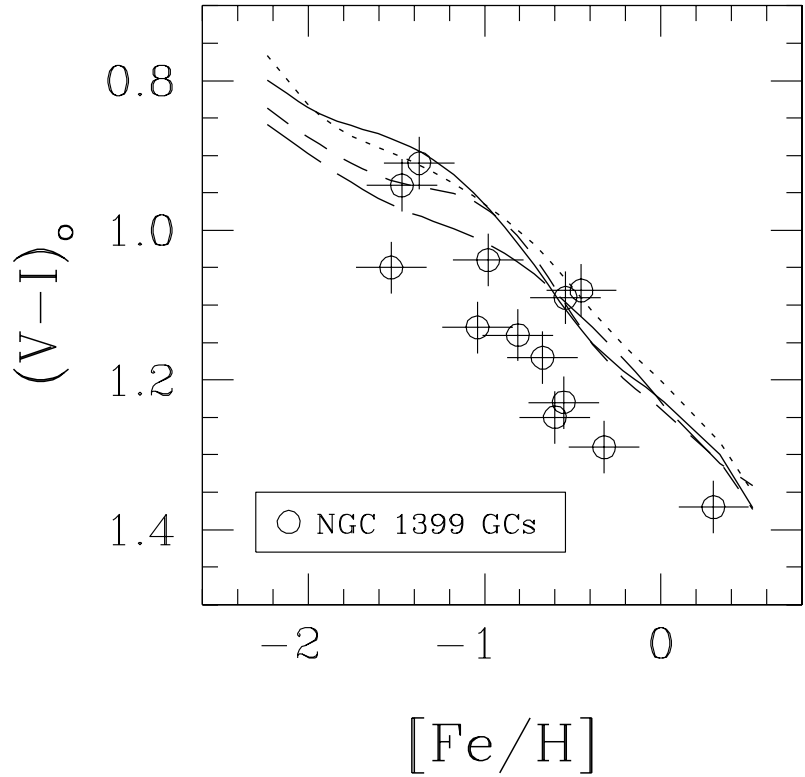}
\caption{The NGC 1399 GCs (photometric data from Kissler-Patig 
et al. 1997, [Fe/H] from Kissler-Patig et al. 1998) are 
compared with our models in the [Fe/H] vs. ($V$ $-$ $I$)$_{o}$ plane. 
Symbols for our models are same as in Fig. 2. 
It appears that there is a systematic offset between the data and our models.} 
\end{figure}

\begin{figure}
\plotone{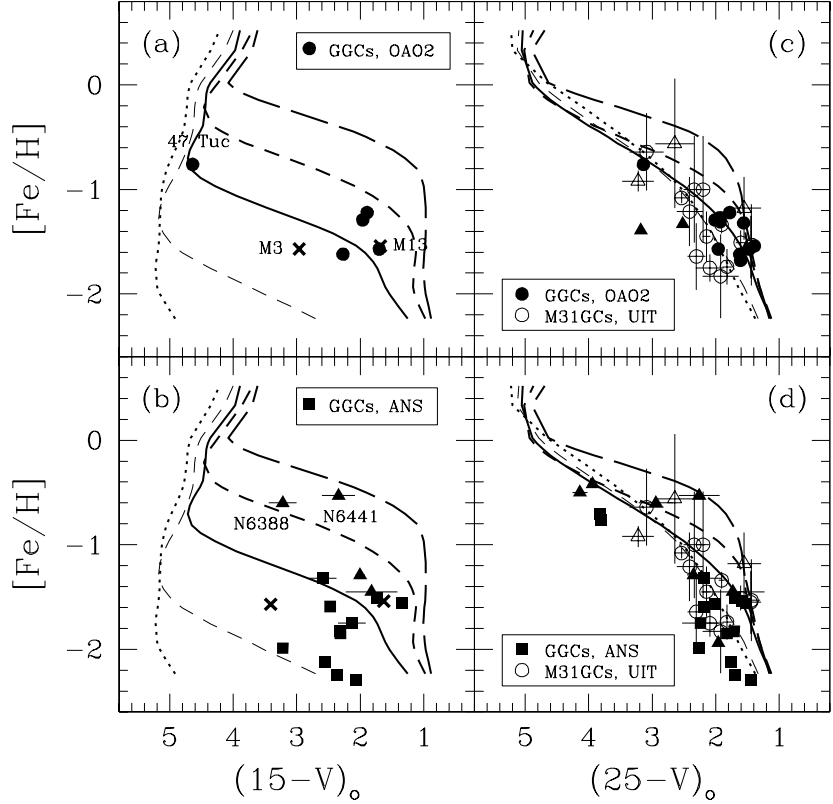}
\caption{
The available UV photometry of Galactic and M31 GCs are compared with 
our models 
in the (15 $-$ $V$)$_{o}$ vs. [Fe/H] [left panel, (a) and (b)] and in the 
(25 $-$ $V$)$_{o}$ vs. [Fe/H] [right panel, (c) and (d)] planes. 
Symbols for our models are same as in Fig. 6. The Galactic 
GCs obtained from {\it OAO 2} are indicated as filled circles and those from 
{\it ANS} are indicated as filled squares, respectively. 
The 2nd parameter Galactic GCs, M3 \& M13, are depicted as crosses
in the left panel. The M31 
GCs observed from {\it UIT} are indicated as open circles. Triangles 
are the data with $E$($B$ $-$ $V$) $>$ 0.2. The Galactic GCs' UV 
photometric data and [Fe/H] are from Dorman et al. (1995) and 
Harris (1996), respectively, and those of M31 GCs are 
from Bohlin et al. (1993) and Barmby et al. (2000), respectively. 
Note that the observations fall within the suggested age range of our 
model predictions (see text).}
\end{figure}

\end{document}